\documentstyle[epsfig,12pt,preprint,tighten,aps]{revtex}
\begin{document}

\draft

\title{\rightline{{\tt June 1999}}
\rightline{{\tt UM-P-99/17}} \rightline{{\tt RCHEP-99/06}} \ \\
Matter-affected neutrino oscillations in ordinary and mirror
stars and their implications for gamma-ray bursts}
\author{Raymond R. Volkas and Yvonne Y. Y. Wong}
\address{School of Physics\\
Research Centre for High Energy Physics\\ The University of
Melbourne\\ Parkville 3052 Australia\\
(r.volkas@physics.unimelb.edu.au, ywong@physics.unimelb.edu.au)}
\maketitle

\begin{abstract}
It has been proposed that the annihilation process $\nu
\overline{\nu} \to e^- e^+ \to \gamma \gamma$ may be responsible
for the generation of gamma-ray bursts (GRBs).  The incipient
neutrino--antineutrino pairs carry virtually all of the
gravitational binding energy available from the central engine.
However, gamma-ray bursters proposed to date are inevitably
surrounded by an excess of baryons, leading to the
``baryon-loading problem''. In the light of growing evidence for
neutrino oscillations, we discuss the implications of
matter-affected oscillations for GRB energetics, and on the
viability of ``mirror'' stars as GRB progenitors.

\end{abstract}

\section{Introduction}

The ongoing quest for a complete understanding of gamma-ray bursts
(GRBs) dates back to their fortuitous discovery in the 1960s
\cite{general}. Over the years, a class of models called the
``fireball model'', which seeks to explain the temporal structure
of the bursts and the non-thermal nature of their spectra, has
emerged \cite{fireball}. In essence, the model consists of a
sudden injection of radiative energy into a compact region with
relatively few baryons ($ \sim 10^{-5} M_{\odot}$), where an
opaque $e^- e^+ \gamma$ plasma is consequently formed. This
``fireball'' expands relativistically, until internal processes
\cite{internal} such as collisions within the now optically thin
outflow trigger the reconversion of its kinetic energy to
radiation. The photons then escape to infinity, to be hailed by us
earthly inhabitants as a GRB. Subsequent deceleration of the
relativistic fireball through interactions with the ambient
interstellar medium gives rise to further emissions at longer
wavelengths:  the so-called afterglow \cite{afterglow}.

Although the discovery of afterglows in the x-ray, optical and/or 
radio domains for several bursts in 1997 \cite{observation} are not
greeted
unanimously by the GRB community as signatures of the classical
fireball model \cite{dar},
the observations have, nonetheless, established firmly for the
first time the cosmological origin of GRBs. Redshifts of $z =
0.835$ and $z = 3.418$ were reported for two of the earliest
observations 
respectively \cite{redshift}, and the recent measurement of $z = 1.600$
in the afterglow of GRB 990123 was performed with unprecedented accuracy
\cite{recent}. The central engines that
power each fireball have not yet been identified, though, at cosmological
distances, these GRB progenitors must be capable of generating
some $10^{51} \to 10^{53}$ ergs of energy in a short period of
time. Prime candidates include binary neutron star merger
\cite{nsnsmerger} and its variants \cite{variants}, and core
collapse \cite{collapses}, in which the concomitant release of
gravitational binding energy is almost entirely in the form of
neutrinos and antineutrinos.  In principle, about one thousandth
of these neutrinos annihilate to form electron--positron pairs,
and ultimately photons \cite{berezinsky} via $\nu \overline{\nu} \to e^-
e^+ \to
\gamma \gamma$, creating a plasma that is the
fireball.\footnote{Of course, the merger/collapse may possess
sufficient rotational energy to power the fireball through the
coupling of its angular momentum to a strong magnetic field
\protect\cite{magnetic}. This, however, is not the topic of this
paper.} The major flaw, however, is that the nascent compact
object is inevitably surrounded by an excess of baryons, which, if
neutrino annihilation is to occur in this environment, would lead
to formation of a nonrelativistic fireball that is inconsistent
with observations.  This is the ``baryon-loading problem''.

Following the recent announcement of very strong evidence for
atmospheric neutrino oscillations at SuperKamiokande
\cite{superk}, a novel solution to the baryon-loading problem that
exploits this simple quantum mechanical phenomenon has been put
forward \cite{kluzniak}. By invoking large amplitude oscillations
between the muon neutrino $\nu_{\mu}$ and a ``sterile'' neutrino
$\nu_s$ with an oscillation length comparable to the width of the
baryonic region, it has been proposed that a neutrino that begins
as a $\nu_{\mu}$ traverses the region largely as a $\nu_s$, and
converts back to $\nu_{\mu}$ upon exit.  The sterile neutrino is,
by definition, inert. Thus annihilation does not take place inside
the baryon-contaminated region, thereby preventing the formation
of a dirty, nonrelativistic fireball.  The ostensible efficiency
of energy deposition is $\sim 10^{-2} \to 10^0$ relative to direct
annihilation in the absence of baryons and oscillations, subject
to the geometry of the GRB progenitor.

Although an attractive idea, the analysis in Ref.\cite{kluzniak}
fails to address two crucial issues: (i) If the central engines
are indeed mergers and/or collapses, there is no reason to assume
that only $\mu$-type neutrinos are (thermally) emitted. Thus {\it
all} neutrino flavours must individually oscillate into a sterile
neutrino to substantially eliminate $\nu \overline{\nu}$
annihilation in the baryonic region.  The conversion of
$\nu_{\mu}$ to $\nu_s$ (and their antiparticles) alone will not
solve the baryon-loading problem. (The only way around this would
be to hypothesise a different type of central engine that produced
$\mu$-type neutrinos entirely from pion decay.)  (ii) Matter
effects may significantly alter the oscillation pattern.
Modifications to the effective neutrino masses and mixings due to
interactions with the medium have, in the past, been studied
extensively in various astrophysical and cosmological contexts. A
notable example is the proposed MSW solution to the solar neutrino
problem \cite{ms}, in which matter effects inside the Sun are
largely responsible for the depletion of the $\nu_e$ flux that
impinges on Earth. Matter-affected oscillations may lay further
claims on the generation of neutrino asymmetries in the early
universe \cite{asymmetry}, and the energetics of and $r$-process
nucleosynthesis  in Type II supernovae \cite{sn,juha}. Gamma-ray
burst progenitors are necessarily dense objects; it is thus our
purpose to reassess the scenario presented in the said analysis in
the light of matter-affected oscillations, and, more generally, to
demonstrate the latter's importance in influencing the energetics
of a neutrino-driven GRB.

Yet, the story does not end here.  It has been suggested that GRBs
may be attributed to the mergers/collapses of ``mirror'' stars
composed of matter that is blind to ordinary interactions
\cite{blinnikov}. The accompanying mirror neutrinos may oscillate
into ordinary neutrinos, whose subsequent annihilation will occur
in regions with few {\it ordinary} baryons, thereby easily
eliminating the baryon-loading problem. In this paper, we shall
also examine this possibility more closely, taking into account
the role of matter effects.

Before proceeding, we should note that one ought to have an open 
mind at this stage as to the mechanism by which GRBs are energised.
Neutrino--antineutrino annihilation may well not be the
sole means (or even {\it a} means) of achieving this end.
Other forms of energy and energy extraction mechanisms, notably
the exploitation of the compact object's spin energy through
coupling to a strong magnetic field \cite{magnetic}, have been
proposed which may be complementary or alternative to the
annihilation process. Be that as it may, neutrino kinetic energy
is certainly a very important source to consider, given its
function as dissipator of gravitational binding energy in mergers
and collapses.  If the neutrino is to play a role, its properties
must be properly understood and its activities incorporated into
prospective GRB models. These form the basis of the present work.

\section{Neutrino oscillations and matter effects}
\label{oscillations}

Neutrino oscillations follow directly from non-degenerate neutrino
masses and non-trivial mixing amongst the flavours. The former
criterion ensures that each propagation eigenstate evolves with a
distinct phase governed by its energy (and thus squared mass by $E
= \sqrt{p^2 + m^2} \simeq p + \frac{m^2}{2 p}$). Subsequent
development of phase differences gives rise to the periodicity of
the oscillation phenomenon, which, for neutrinos of momentum $p$
in vacuum, is determined by the ratio $2 \pi \frac{2 p}{\Delta m^2_{ij}}$,
 where
$\Delta m^2_{ij} = m^2_i - m^2_j$ is the squared mass difference
between the $i$th and $j$th mass (propagation) eigenstates. The
oscillation amplitude scales with the amount of mixing between the
states. For a two-neutrino system, this is characterised by one
mixing angle $\theta$, such that
\begin{eqnarray}
| \nu_{\alpha} \rangle & = & \cos \theta | \nu_1 \rangle + \sin
\theta | \nu_2 \rangle, \nonumber \\ | \nu_{\beta} \rangle & = & -
\sin \theta | \nu_1 \rangle + \cos \theta | \nu_2 \rangle,
\end{eqnarray}
where $\nu_1$ and $\nu_2$ are the mass eigenstates, and the
subscripts $\alpha$ and $\beta$ label two different flavour
eigenstates respectively. By the $CPT$ theorem, the same
oscillation parameters govern the flavour evolution of both
neutrino and antineutrino systems in vacuum.

In the presence of matter, the neutrino gains an effective mass
from interacting with the ambience \cite{wolfenstein}. The nature
of the gain
--- its magnitude and sign --- is subject to the density
of the medium and the interaction channels that are available
therein. Thus two oscillating neutrino flavours that interact
differently with the environment will develop between them a phase
difference that is dissimilar to its vacuum counterpart, thereby
modifying the oscillation length and, though somewhat less
obviously from our qualitative discussion, the oscillation
amplitude.  For a two-neutrino system\footnote{Hereafter, we shall
consider only two-neutrino systems.} in a {\it uniform} medium,
the probability that $\nu_{\alpha}$ will oscillate to
$\nu_{\beta}$, where $\alpha \neq \beta$, at time $t$ is given by
\cite{kuo}
\begin{equation}
\label{probability}
P(\alpha \to \beta, t) = \sin^2 2 \theta_{\text{eff}}
        \sin^2 \frac{\pi t }{\lambda_{\text{eff}}},
\end{equation}
where the quantity
$\lambda_{\text{eff}} = 2 \pi \frac{2 E}{\Delta m^2_{\text{eff}}}$
is the effective oscillation length, and
\begin{eqnarray}
\label{effective} \Delta m^2_{\text{eff}} = \Delta m^2_{\alpha \beta}
\sqrt{ \left( \frac{2 E V_{\alpha \beta}}{\Delta m^2_{\alpha \beta}}
- \cos 2 \theta \right)^2 + \sin^2 2 \theta}, \nonumber \\
\sin^2 2 \theta_{\text{eff}} =
    \frac{\sin^2 2 \theta}{\left( \frac{2 E V_{\alpha \beta}}
        {\Delta m^2_{\alpha \beta}}
    - \cos 2 \theta \right)^2 + \sin^2 2 \theta},
\end{eqnarray}
with $V_{\alpha \beta} = \Phi_{\alpha} - \Phi_{\beta}$, where
$\Phi_{\alpha}$ ($\Phi_{\beta}$) is the matter potential for
$\nu_{\alpha}$ ($\nu_{\beta}$), and we have used $E \simeq p$.
Note that for clarity, all squared mass differences will now carry
the subscripts $\alpha \beta$ (denoting flavours) such that
$\Delta m^2_{\mu \tau}$, for example, corresponds to the squared
mass difference between the two mass eigenstates relevant for the
$\nu_{\mu} \leftrightarrow \nu_{\tau}$ system.

A typical celestial medium is an electrically neutral concoction
of electrons/positrons and nucleons (both bound and free).  Thus a
$\nu_e$ propagating therein has both charged and neutral current
interactions, while $\nu_{\mu}$ and $\nu_{\tau}$ have only the
latter, and $\nu_s$ has none.  To the lowest order in $G_F$, their
respective matter potentials are
\begin{eqnarray}
\label{potentials}
\Phi_e & = & \sqrt{2} G_F \left( N_e - \frac{1}{2}N_n \right)
    = \frac{G_F}{\sqrt{2}} \frac{\rho}{m_N}
    \left( 3Y_e - 1 \right), \nonumber \\
\Phi_{\mu}& = & \Phi_{\tau} = - \frac{G_F}{\sqrt{2}} N_n = \frac{G_F}{\sqrt{2}}
    \frac{\rho}{m_N} \left(Y_e - 1 \right), \nonumber   \\
\Phi_s & = & 0,
\end{eqnarray}
where $G_F$ is the Fermi constant, $N_e$ denotes the electron
minus positron number density, $N_n$ the neutron number density,
$\rho$ the nucleon density, $m_N$ the nucleon mass, and $Y_e$ the
number of electrons per nucleon. Note that for antineutrinos,
$\Phi_{\overline{\alpha}} = - \Phi_{\alpha}$, such that a
$\overline{\nu}_{\alpha} \leftrightarrow \overline{\nu}_{\beta}$
system receives modifications to its effective oscillating
parameters generally unlike those for a $\nu_{\alpha}
\leftrightarrow \nu_{\beta}$ system in an identical medium.

\section{Oscillations in GRB progenitors}

Accompanying a merger/collapse event is the copious production of
$\nu_e,\; \nu_{\mu}$ and $\nu_{\tau}$ and their antiparticles,
with mean energies ranging from $\sim 10$ to $\sim 30$ MeV.  In
this section, we shall examine the oscillations of these
``active'' neutrinos with (i) ``sterile'' neutrinos, and (ii)
amongst themselves in GRB progenitors {\it per se}, adhering
strictly only to laboratory bounds on the oscillation parameters.
Constraints arising from cosmological circumstances such as big
bang nucleosynthesis (BBN) and closure will be noted at the
appropriate points.

\subsection{Active--sterile oscillations}

We shall suppose here that each active flavour mixes with a
sterile neutrino. The consideration of sterile neutrinos is well
motivated for a number of reasons:
\begin{enumerate}
\item Right-handed neutrinos, which are sterile with respect to
ordinary weak interactions, are necessary for a complete
correspondence between lepton and quark degrees of freedom in the
standard model of particle physics.
\item A particular class of
light, effectively sterile fermions called ``mirror neutrinos''
arises if Improper Lorentz Transformations are retained as exact
symmetries of Nature (see later).
\item Phenomenologically, they
are strongly advocated through the need to resolve the apparent
conflict between the three neutrino anomalies --- solar
\cite{solar}, atmospheric \cite{atmospheric} and LSND \cite{lsnd}
--- and the measured width of the $Z^0$ boson. The former in its
entirety calls for an oscillation solution requiring at least four
neutrinos, while the latter constrains the number of light active
flavours to three.
\end{enumerate}
Experiments performed thus far do not preclude the existence of
yet more sterile species.  Indeed, it is theoretically quite
natural for the number of light sterile flavours to equal the
number of quark and lepton generations, viz. three; this is our
assumption for the rest of the paper.  Furthermore, if the
``sterile'' flavours are identified with mirror neutrinos as in
point 2 above, then they must come in triplicate.  Thus our
analysis here will also set the stage for the study of mirror
stars in the next section.

\subsubsection{Large mixing angle}

Suppose that each active flavour $\nu_{\alpha}$ exhibits large
vacuum mixing with a sterile ``partner'' $\nu_{\alpha}'$, that is,
$\cos 2 \theta \approx 0$.\footnote{From here onwards, the symbol
$\nu_s$ shall denote a generic sterile neutrino, while
$\nu_{\alpha}'$ is taken to mean the assigned sterile partner of
$\nu_{\alpha}$, one for each of $\nu_e$, $\nu_{\mu}$ and
$\nu_{\tau}$.} This is a most natural consequence from the
perspective of model building, arising from the general
Dirac--Majorana mass matrix for each generation of neutrinos,
\begin{equation}
\left[ \begin{array}{cc}
            \overline{\nu}_L & \overline{(\nu_R)^c} \end{array}
            \right]
\left( \begin{array}{cc}
            0 & m \\
            m & M \end{array} \right)
\left[ \begin{array}{c}
            (\nu_L)^c \\
            \nu_R \end{array} \right],
\end{equation}
where the Dirac mass $m$ is taken to be much larger than the
right-handed neutrino Majorana mass $M$.\footnote{Note that the
zero in the top-left corner of the mass matrix is enforced by
electroweak gauge invariance in the absence of weak-isospin
triplet Higgs bosons.} Upon diagonalisation, the two resulting
pseudo-Dirac neutrinos, one of which we identify as the
right-handed sterile neutrino, are essentially maximally mixed
\cite{pseudoDirac}. Pairwise maximal mixing also arises in the
mirror matter model.  Indeed, large amplitude pairwise
oscillations of the active flavours into distinct sterile states
are {\it a priori} necessary if {\it all} active $\nu
\overline{\nu}$ annihilation in the baryon-contaminated mantle is
to be prevented.

For this case of (almost) maximal mixing, matter effects are
virtually identical for both neutrinos and antineutrinos by Eq.\
(\ref{effective}). From an inspection of the same equation, we
identify two regions of interest:
\begin{equation}
\label{condition}
 \frac{\Delta m^2_{\alpha \alpha'}}{2 E}
\quad \begin{array}{c}    \ll \\
                    \stackrel{>}{\sim} \end{array} \quad
\left| V_{\alpha \alpha'} \right|,
\end{equation}
where we have used $\sin 2 \theta \approx 1$.  These shall be labelled as the
first and second conditions respectively.
For convenience, we rewrite
Eq.\ (\ref{condition}) in more accessible units,
\begin{eqnarray}
\label{accessible} \frac{\Delta m^2_{ee'}}{E} \quad &
\begin{array}{c} \ll \\
                   \stackrel{>}{\sim}\end{array} \quad  &
    \left| 760 \left( \frac{\rho}{10^{10} \; \text{g} \,
    \text{cm}^{-3}} \right) \left( 3 Y_e - 1 \right) \right| \;
    \frac{\text{eV}^2}{\text{MeV}} \quad \quad \text{for} \; \nu_e
    \leftrightarrow \nu_e', \nonumber\\
\frac{\Delta m^2_{\mu \mu',\, \tau \tau'}}{E} \quad  &
\begin{array}{c} \ll
\\
                \stackrel{>}{\sim} \end{array} \quad &
    \left| 760 \left( \frac{\rho}{10^{10}\; \text{g}\, \text{cm}^{-3}}
    \right) \left(Y_e - 1 \right) \right| \;
    \frac{\text{eV}^2}{\text{MeV}} \quad \quad \text{for}\;
    \nu_{\mu,\, \tau} \leftrightarrow \nu_{\mu,\, \tau}',
\end{eqnarray}
where the symbols are defined as for Eq.\
(\ref{potentials}).\footnote{The presence of neutrinos contributes
to the matter potentials in  Eq.\ (\protect\ref{potentials}).
However, we do not expect such a contribution to have too serious
a consequence since the effective neutrino number density is
generally small except near the neutrino emitting surface. In any
case, the exclusion of the neutrino background should not alter
the qualitative aspect of the present work.}

In accordance with Eq.\ (\ref{effective}), the first condition
corresponds to $\sin 2 \theta_{\text{eff}} \to 0$, implying that
oscillations are strongly suppressed.  Typically, the density of
the resultant disk in a binary neutron star merger is at least
$\rho \sim 10^{9}\; \text{g}\, \text{cm}^{-3}$ with $Y_e \sim 0.02
\to 0.1$ \cite{nsnsmerger}, while that of the surrounding mantle
in a collapse event is expected to be no less than $\rho \sim
10^{6}\; \text{g}\, \text{cm}^{-3}$ with $Y_e \sim 0.2 \to
0.5$.\footnote{These numbers correspond to the density at $r \sim
300$ km and the number of electrons per nucleon, respectively, in
a Type II supernova at $\sim 0.6$ s post bounce
\protect\cite{wilson}.} The SuperKamiokande results put the
squared mass difference for $\nu_{\mu} \leftrightarrow \nu_x$,
where $\nu_x$ is some as yet unidentified neutrino, at $10^{-3}
\stackrel{<}{\sim} \Delta m^2 / \text{eV}^2 \stackrel{<}{\sim}
10^{-2}$\cite{superk}.   The corresponding upper bound for maximal
$\nu_{e} \leftrightarrow \nu_x$ mixing is currently $\sim
10^{-3}\; \text{eV}^2$ \cite{chooz}. It follows from Eq.
(\ref{accessible}) that the average $10 \to 30$ MeV $\nu_e$ and
$\nu_{\mu}$ have virtually no chance of oscillating into a sterile
species inside the baryonic region. Oscillations become more
suppressed with the increase of neutrino energy. The formation of
a dirty fireball therefore cannot be avoided with the introduction
of $\nu_{\mu} \leftrightarrow \nu_{\mu}'$ or $\nu_e
\leftrightarrow \nu_e'$ oscillations, which argues against the
scenario of Ref.\cite{kluzniak}.

The second condition in Eq.\ (\ref{accessible}) entails
vacuum-like maximal oscillations. Taking the density of the
neutrino emitting surface to be $\rho \sim 10^{11}\; \text{g}\,
\text{cm}^{-3}$, it is clear from Eq.\ (\ref{accessible}) that
matter effects remain unimportant in all or part of the baryonic
region only for sub-keV $\nu_e$ and $\nu_{\mu}$.  But these
neutrinos are of little consequence; apart from inabundance, their
energies are well below the threshold for $e^- e^+$ pair
production.

At this stage, the acute reader would have noticed that in the
case of core collapse, complete cancellation of matter effects for
a $\nu_e \leftrightarrow \nu_e'$ system arises when $Y_e \approx
0.33$ by Eq.\ (\ref{accessible}), where the effective mixing is
temporarily vacuum-like and thus maximal.  Substantial $\nu_e'$'s
may be generated if such cancellation persists (approximately)
over a distance comparable to the effective oscillation length of
the system (i.e., the adiabatic condition
--- see later). Supposing that $\Delta m^2_{ee'} \sim 10^{-3}\;
\text{eV}^2$ and $E \sim 10$ MeV, Eq.\ (\ref{accessible}) demands
the change in $Y_e$ in this region to be less than $\sim 10^{-4}$
even at a fixed density as low as $\rho \sim10^{6}\; \text{g}\,
\text{cm}^{-3}$. Holding $Y_e$ constant at, say, $0.33 +
10^{-10}$, the same equation requires that any deviation in
density to be $< 10^{5}\; \text{g}\, \text{cm}^{-3}$. However,
given the dramatic rise and fall of $Y_e$ and the density
respectively in a mere few hundred kilometres, and that the
oscillation length for the system concerned is $\sim 25$ km, the
said conditions are unlikely to be satisfied across a region
comparable to the latter.  The production of $\nu_e'$'s is again
suppressed, albeit by a different mechanism.

On the other hand, no laboratory upper bound on $\Delta m^2$
exists for maximal $\nu_{\tau} \leftrightarrow \nu_{\tau}'$
mixing.  There are cosmological constraints from closure and big
bang nucleosynthesis. The former yields an upper bound of about
$40 \to 100$ eV for long lived neutrinos.\footnote{Purported BBN
constraints must be interpreted with care because important
loopholes frequently exist. However, one can safely say that a
maximally mixed active--sterile pair with a $\Delta m^2$ value in
the range to be considered is disfavoured by BBN.}  Supposing
$\nu_{\tau}$ to be much lighter than $\nu_{\tau}'$ or vice versa,
this condition effectively sets an upper limit of $\sim 10^4\;
\text{eV}^2$ on the squared mass difference.  It transpires that
if one pushes $\Delta m^2_{\tau \tau'}$ to the extreme, it is in
fact possible to attain vacuum-like maximal $\nu_{\tau}$
oscillations with its sterile partner even at a density of $\rho
\sim 10^{10}\; \text{g}\, \text{cm}^{-3}$, according to Eq.\
(\ref{accessible}). The cost of an increased $\Delta m^2$,
however, is the simultaneous shortening of the oscillation length.
If the latter is to be comparable to the size of the baryonic
region $R$ and approximately maximal mixing is to be maintained
throughout, then by Eqs.\ (\ref{effective}) and (\ref{condition}),
the following condition must hold:
\begin{equation}
2 \pi \stackrel{>}{\sim} \left| V_{\tau \tau'}^{\text{max}} R \right|,
\end{equation}
or equivalently,
\begin{equation}
\label{noway}
     \left| \left( \frac{R}{\text{km}} \right) \left(
\frac{\rho_{\text{max}}}{10^{10}\;
\text{g}\, \text{cm}^{-3}} \right) \left(Y_e - 1 \right) \right|
\stackrel{<}{\sim}  3.3 \times 10^{-6},
\end{equation}
where ``max'' denotes maximum beyond the neutrinosphere. Given that
the average neutrino traverses a few kilometres of baryonic matter
in a merger, not to mention the extent of the mantle in a collapse
event, the reader can verify that Eq.\ (\ref{noway}) cannot be
satisfied in any realistic GRB progenitor. Instead, the system
undergoes rapid oscillations, as implied by its comparatively
short oscillation length, quickly becoming, on average, an equal
mixture of $\nu_{\tau}$ and $\nu_{\tau}'$ (and their
antiparticles) by Eq.\ (\ref{probability}).   This scenario,
however, is deemed unlikely for cosmological reasons mentioned
earlier. But if some oscillations were to occur (perhaps with a
smaller $\Delta m^2$), the $\nu_{\tau}$ and
$\overline{\nu}_{\tau}$ intensities at $r$ would be, respectively,
effectively halved such that $\nu_{\tau} \overline{\nu}_{\tau}$
annihilation would still take place inside the baryonic region,
but at a quarter of the standard rate per unit volume.  Assuming
that all active flavours contribute equally to annihilation in the
absence of oscillations, the total energy deposition rate in our
case is expected to suffer at worst a $25$ \% decrease.

\subsubsection{Small mixing angle}

An interesting effect arises for propagation in a medium of
monotonically varying density.  A level-crossing occurs when the
neutrinos traverse a region in which the effective masses are
virtually degenerate, i.e., where the resonance condition
\begin{equation}
\label{resonance}
 2 E V_{\alpha \beta} =  \Delta m^2_{\alpha \beta} \cos 2 \theta,
\end{equation}
is satisfied.   Provided that the matter density is changing
sufficiently slowly, a $\nu_{\alpha}$ entering the resonance will
emerge as a $\nu_{\beta}$, where $\alpha \neq \beta$, and vice
versa. This is the Mikheyev--Smirnov--Wolfenstein (MSW)
effect \cite{ms,wolfenstein} and is particularly prominent for
$\sin 2 \theta \sim 0$. The conversion efficiency depends on the
ratio of the physical width of the resonance region to the corresponding
effective oscillation
length of the system, or equivalently,
\begin{equation}
\gamma \equiv \left. \frac{\left( \frac{\Delta m^2_{\alpha
\beta}}{2 E} \sin 2 \theta \right)^2}{\left| \frac{d V_{\alpha
\beta}}{d r}
        \right|} \right|_{\text{res}},
\end{equation}
where $\frac{d V_{\alpha \beta}}{dr}$ is the rate of change of the matter
potential along the neutrino's path.
The larger the ratio (otherwise known as the adiabaticity
parameter), the more effective the transformation.

 Our interest in
the case of small vacuum mixing lies in the possible existence of
such a resonance within the baryonic region. If Eq.\
(\ref{resonance}) is satisfied therein and the adiabaticity
parameter $\gamma$ is sufficiently large, the ensuing conversion
of all active neutrinos to sterile species means that, beyond the
resonance, no neutrinos are available for annihilation. This loss
of energy is practically irretrievable, unless a second resonance
exists through which steriles reconvert to actives.\footnote{This
situation is in fact not as contrived as it first seems. A double
$\nu_e \leftrightarrow \nu_s$ resonance has been shown to exist in
the post bounce hot bubble in a Type II supernova
\protect\cite{juha}.  However, this possibility will not be dealt
with here, owing to its extreme dependence on the density profile
of the progenitor; the spatial distribution of baryons in a Type
II supernova is perhaps not representative of those in core
collapse scenarios in general.} The reduction in the total energy
deposition rate hinges on the location of the resonance, since the
$\nu \overline{\nu}$ annihilation rate per unit volume $q$
generally has an $r$-dependence, such as $q \propto r^{-8}$ for
spherical geometry \cite{rate}. We leave this calculation for the
interested numerical modeller.

We now explore the parameter space in which resonant conversion to
sterile neutrinos would significantly decrease the energy
deposition by $\nu \overline{\nu}$ annihilation, by calculating
the constraints on the oscillation parameters required for the
prevention of energy loss via this mechanism.  Consider a binary
neutron star merger, and let us suppose that each active species
exhibits small mixing only with its sterile partner. Assuming that
each $\nu_{\alpha}'$ is lighter than its active counterpart, the
extremely low value of $Y_e$ in this environment means that the
resonance condition can only be satisfied by antineutrino systems,
as indicated by Eq.\ (\ref{potentials}). Given the relevant
densities, $\rho \sim 10^{9} \to 10^{11} \; \text{g}\,
\text{cm}^{-3}$, the average $20 \to 30$ MeV (anti)neutrino will
undergo resonant conversion if the squared mass difference of the
oscillating system happens to lie in the approximate range
\begin{equation}
\label{mass}
10^{3} \stackrel{<}{\sim} \Delta m^2 / \text{eV}^2
\stackrel{<}{\sim} 10^5.
\end{equation}
Furthermore, by holding the quantity $Y_e$ constant, we rewrite
the adiabaticity parameter in more civilised units,
\begin{equation}
\label{ad}
\gamma = \left. \frac{3}{\eta} \left[ \left(
\frac{\Delta m^2}{\text{eV}^2} \right) \left( \frac{\text{MeV}}{E}
\right) \sin 2 \theta \right]^2 \left( \frac{10^{10}\;\text{g}\,
\text{cm}^{-3}\, \text{km}^{-1}}{\left| \frac{d \rho}{d r}
\right|} \right) \right|_{\text{res}},
\end{equation}
where $\eta  =  1 - 3 Y_e$ and $1 - Y_e$ for $\overline{\nu}_e
\leftrightarrow \overline{\nu}_e'$ and $\overline{\nu}_{\mu,\,
\tau} \leftrightarrow \overline{\nu}_{\mu,\,\tau}'$ respectively.
One may reasonably expect the density gradient $\frac{d \rho}{dr}$
to be some undoubtedly highly model-dependent function of $r$. For
our crude analysis, we make the approximation
\begin{equation}
\label{gradient}
\frac{d \rho}{dr} \approx \frac{\rho_{\text{max}} -
\rho_{\text{min}}}{R} \sim \frac{10^{11}\; \text{g}\,
\text{cm}^{-3}}{10\; \text{km}} = 10^{10}\; \text{g}\,
\text{cm}^{-3}\, \text{km}^{-1}.
\end{equation}
If we demand $\gamma \ll 1$ such that resonant conversion to
$\overline{\nu}_s$ is ``non-adiabatic'' and thus inefficient, then
by Eqs.\ (\ref{ad}) and (\ref{gradient}) together with $Y_e =
0.05$, the following approximate constraints on the vacuum mixing angle
are obtained,
\begin{equation}
\sin^2 2 \theta \ll 10^{-4} \to 10^{-8},
\end{equation}
for the range of squared mass differences in Eq.\
(\ref{mass}), where we have taken the neutrino energy to be the
average $20 \to 30$ MeV intrinsic to binary neutron star mergers.

\subsection{Active--active oscillations}

In the following, we briefly examine the consequences of mixing
amongst the active flavours.

Oscillations between $\nu_{\mu}$ and $\nu_{\tau}$ are not affected
by the presence of matter, since they interact similarly with
ordinary matter.  For the same reason, these thermal neutrinos are
produced with identical energy spectra and are therefore of little
interest from the perspective of $\nu_{\mu} \leftrightarrow
\nu_{\tau}$ oscillations.

Contrastingly, the $\nu_e \leftrightarrow \nu_{\mu,\, \tau}$
system may experience resonant conversion, given the correct
oscillation parameters. Since $\nu_e$'s are more abundant, this
implies a possible decrease in the $\nu_e$ flux beyond the
resonance. However, as the more energetic $\nu_{\mu,\,\tau}$'s are
simultaneously converted to $\nu_e$'s, one may reasonably expect
the reduction in flux to be compensated for by a harder spectrum.
Similarly, the increase in $\nu_{\mu,\,\tau}$ flux is accompanied
by a softening of the spectrum. Thus, summing over all flavours,
the energy deposition rate due to $\nu \overline{\nu}$
annihilation should, to a first approximation, exhibit minimal
difference from the no-oscillation case.

\section{Mirror stars}

The concept of a mirror world was introduced as a means to retain
parity and time-reversal transformations (Improper Lorentz
Transformations) as exact symmetries of Nature.  In essence, the
content of the Standard Model of particle physics is enlarged to
include a mirror sector such that every ordinary particle is
partnered with a mirror image differing only in its handedness.
The resulting theory has been called the Exact Parity Model
\cite{epm} (see also Ref.\cite{mohapatra} for a different model). These
particles participate in mirror interactions
identical in nature to ordinary processes, but are inert with
respect to the ordinary strong, electromagnetic and weak forces.
Thus the mirror world evolves as we do, complete with stellar
mergers and collapses, its only link to the ordinary world being
through gravitational coupling, and the mixing of colourless and
electrically neutral ordinary--mirror partners. If neutrinos have
non-degenerate masses, maximal ordinary--mirror neutrino
oscillations are a necessary consequence of the underlying exact
parity symmetry. Interestingly, the maximal mixing of $\nu_e$ with
its mirror partner can solve the solar neutrino problem, while the
maximal mixing of $\nu_\mu$ with {\it its} mirror partner can
solve the atmospheric neutrino problem \cite{solution}.

Recently, Blinnikov has proposed that the central engines of GRBs
may be cataclysmic astrophysical events involving mirror stars
\cite{blinnikov}. We now examine the implications of
matter-affected neutrino oscillations for this proposal.

Mirror neutrinos emitted in a mirror merger/collapse must traverse
a region of excess mirror baryons and suffer the same matter
effects as do their ordinary counterparts.  Thus interactions
between mirror neutrinos and the mirror ambience are equally well
described by the matter potentials written down earlier in Eq.\
(\ref{potentials}), save for a change of labels
--- $\alpha$ becomes $\alpha'$, where the primed symbol now denotes a
mirror particle. In this environment, our ordinary $\nu_e$,
$\nu_{\mu}$ and $\nu_{\tau}$ are effectively what were previously
labelled as sterile neutrinos.

Given that the $\nu \overline{\nu}$ annihilation rate per unit
volume generally decreases with $r$, in order to channel as much
energy as possible towards the generation of an {\it ordinary}
GRB, rapid maximal ordinary--mirror oscillations for both
neutrinos and antineutrinos throughout the progenitor is desired.
However, as suggested by results from the previous section, this
situation cannot be realised by the $\nu_e \leftrightarrow \nu_e'$
and $\nu_{\mu} \leftrightarrow \nu_{\mu}'$ systems for which
oscillations are highly suppressed at the nominal densities.
Pushing the squared mass differences to their respective upper
limits, substantial mixing is possible at densities lower than
$\rho \sim 10^{3} \to 10^{4}\; \text{g}\, \text{cm}^{-3}$ by Eq.\
(\ref{accessible}). But the annihilation of ordinary neutrinos
will now take place at large distances where the rate is rendered
insignificant by geometric factors.  As an illustration, suppose
that the progenitor is spherical and that ordinary $\nu_e$ and
$\overline{\nu}_e$ are available in large quantities only at $r
> r_0$ where mixing is not suppressed.
We estimate the efficiency of the energy deposition due to $\nu_e
\overline{\nu}_e$ annihilation to be
\begin{equation}
\label{illustration} \frac{\dot{Q}_{e
\overline{e}}}{\dot{Q}^{\text{ord}}_{e \overline{e}}} \simeq
\frac{L_e L_{\overline{e}} \int^{\infty}_{r_0} r^{-8} r^2 dr}{
L^{\text{ord}}_{e} L^{\text{ord}}_{\overline{e}}
\int^{\infty}_{r_{\nu}} r^{-8} r^2 dr}   = \frac{L_e
L_{\overline{e}}}{L^{\text{ord}}_e L^{\text{ord}}_{\overline{e}}}
\left( \frac{r_0}{r_{\nu}} \right)^{-5},
\end{equation}
where $\dot{Q}$ is the integrated energy deposition rate, $L_e$
and $L_{\overline{e}}$ are the effective luminosities of $\nu_e$
and $\overline{\nu}_e$ respectively, $r_{\nu}$ the radius of the
emitting surface, and the subscript ``ord'' denotes ordinary.
Equation (\ref{illustration}) clearly demonstrates that a distance
as small as $r_0 \approx 4 r_{\nu}$ is enough to produce at least
a thousand-fold decrease in the efficiency (since $L_{\alpha} \leq
L^{\text{ord}}_{\alpha}$).  Thus ordinary $\nu_e \overline{\nu}_e$
and $\nu_{\mu} \overline{\nu}_{\mu}$ annihilation in a mirror
event may be safely ignored.

 Conversely, the $\nu_{\tau}
\leftrightarrow \nu_{\tau}'$ system may at least partially fulfil
the aforementioned requirements, if $\Delta m^2_{\tau \tau'}$ is
sufficiently large for maximal mixing to be attained not too far
from the neutrinosphere (with the usual caveats regarding possible
cosmological constraints understood).  Be this the case, rapid
maximal oscillations will lead to the effective generation of a
$\nu_{\tau}$ and a $\overline{\nu}_{\tau}$ flux, each with a
luminosity equal to half of that expected from an ordinary
merger/collapse. This implies that the energy deposition rate per
unit volume at $r$ is a factor of four smaller than that due to
$\nu_{\tau} \overline{\nu}_{\tau}$ annihilation alone in an
ordinary event. In the standard picture, all three active flavours
contribute roughly equal amounts of energy towards the burst. It
follows that the total annihilation rate per unit volume at $r$
must be some ten times less than the ordinary rate.  Furthermore,
that ordinary annihilation only takes place at $r > r_0$
introduces a geometric reduction factor. Thus, assuming spherical
geometry, we estimate the overall efficiency of energy deposition
to be
\begin{equation}
\frac{\dot{Q}_{\text{total}}}{\dot{Q}^{\text{ord}}_{\text{total}}}
\simeq \frac{\dot{Q}_{\tau \overline{\tau}}}{\sum_{\alpha = e,\,
\mu,\, \tau} \dot{Q}^{\text{ord}}_{\alpha \overline{\alpha}}}
\simeq \frac{L_{\tau} L_{\overline{\tau}}}{\sum_{\alpha = e,\,\mu,
\, \tau} L^{\text{ord}}_{\alpha}
L^{\text{ord}}_{\overline{\alpha}}} \left( \frac{r_0}{r_{\nu}}
\right)^{-5} \simeq  \frac{1}{10}\left( \frac{r_0}{r_{\nu}}
\right)^{-5},
\end{equation}
where the assumption $r_{\nu_e} \approx r_{\overline{\nu}_e}
\approx \cdots \approx r_{\nu_{\tau}} \equiv r_{\nu}$ is implicit.
As an illustration, the matter density in a core collapse is such
that a $\nu_{\tau} \leftrightarrow \nu_{\tau}'$ system with
$\Delta m^2 \sim 10\; \text{eV}^2$ may enjoy maximal mixing beyond
$r_0 \sim 2 r_{\nu}$.\footnote{These numbers are inferred from
Figure 1b in Ref.\ \protect\cite{juha} for a Type II supernova at
$\sim 6$ s post bounce.}
 Thus a GRB generated by such  a mirror event must be
approximately $300$ times less energetic than one produced by an
equivalent event in the ordinary world.  The reward, however, is
that the baryon-loading problem is virtually eliminated.

We shall not consider small angle resonant conversion of mirror to
ordinary neutrinos since this process is generally not
simultaneously available for both neutrinos and antineutrinos.
Ordinary annihilation necessarily requires the presence of both
$\nu$ and $\overline{\nu}$.  Thus small mixing between mirror
partners alone will not lead to the production of ordinary GRBs in
mirror events.

\section{Summary and conclusion}

Matter-affected neutrino oscillations in GRB progenitors are
studied in this paper.  For simplicity, all oscillation schemes
examined are essentially independent two-neutrino systems.  It is
found that oscillations amongst the ordinary, active flavours ---
$\nu_e$, $\nu_{\mu}$ and $\nu_{\tau}$ --- have minimal effects on
the energetics of the burst. Maximal $\nu_e$ and $\nu_{\mu}$
oscillations with their respective sterile partners are also
expected to be of little consequence. Contrastingly, if
$\nu_{\tau}$ is allowed to oscillate maximally to its sterile
partner with a squared mass difference $\Delta m^2
\stackrel{>}{\sim} 10^4\; \text{eV}^2$, the energy available for
the ultimate GRB may suffer a $25$ \% decrease.  However,
reconciliation with constraints imposed by cosmological closure
and big bang nucleosynthesis renders this option unlikely.

In the small mixing angle regime, the possible existence of an MSW
resonance in the baryonic region implies a generally irretrievable
loss of energy beyond the resonance in the form of sterile
neutrinos. By demanding minimal loss, we are able to determine
some crude constraints on the oscillation parameters.  These
can be found in the appropriate section in the paper.

Contrary to earlier claims, matter effects alter the oscillation pattern
in such a way that the ``temporary'' conversion to $\nu_s$ as a means to
bypass the baryonic region
cannot be achieved in any realistic GRB progenitor.  The fireball will
remain as dirty as dictated by the merger/collapse.

The suppression of mirror to ordinary neutrino oscillations by matter 
effects also argues against the viability of mirror mergers/collapses as
{\it ordinary} GRB progenitors.  Even the most efficient
$\nu_{\tau} \leftrightarrow \nu_{\tau}'$ maximal oscillations with
$\Delta m^2 \stackrel{>}{\sim} 10^4\; \text{eV}^2$ would lead to
some factor of ten decrease in the energy of the resultant burst
relative to that generated by an equivalent event in the ordinary
world.  Further deterioration inevitably follows, at least in the
case of spherical geometry, any decrease in the squared mass
difference.   Ultimately, the central mirror engine will perhaps
need to be a few hundred (or more) times more energetic than its
ordinary counterpart if it is to produce an ordinary GRB that is
compatible in energy with observations. However, with the
guaranteed elimination of the baryon-loading problem, this remains
an option.

We stress at this point that the study of matter-affected
oscillations is highly model-dependent --- the word ``model''
referring to both the GRB and the neutrino model. Analyses of
two-neutrino systems merely serve to illustrate some of the
possible effects.  But most importantly, we wish to emphasise the
necessity to consider matter effects on the oscillation pattern,
if neutrinos are to be the means of energy transportation in any
GRB progenitor. At this stage, there is no clear evidence for the
correct GRB or neutrino model. Hopefully, with new neutrino
experiments underway, the latter will be at least partially
resolved in the not too distant future.

\acknowledgments{This work was supported in part by the Australian
Research Council and in part by the Commonwealth of Australia's
postgraduate award scheme.}


\begin{thebibliography}{99}

\bibitem{general}
For a review on gamma-ray bursts, see T.\ Piran, astro-ph/9810256;
P.\ M\'{e}sz\'{a}ros, astro-ph/9904038.

\bibitem{fireball}
B.\ Paczy\'{n}ski, Astrophys.\ J.\ Lett.\ {\bf 308}, L43 (1986);
J.\ Goodman, {\it ibid.} {\bf 308}, L47 (1986); A.\ Shemi and T.\
Piran, {\it ibid.} {\bf 365}, L55 (1990); J.\ H.\ Krolik and E.\
A.\ Pier, Astrophys.\ J.\ {\bf 373}, 277 (1991).

\bibitem{internal}
R.\ Narayan, B.\ Paczy\'{n}ski and T.\ Piran, Astrophys.\ J.\
Lett.\ {\bf 395}, L83 (1992); M.\ J.\ Rees and P.\
M\'{e}sz\'{a}ros, {\it ibid.} {\bf 430}, L93, (1994); B.\
Paczy\'{n}ski and G.\ Xu, Astrophys.\ J.\ {\bf 427}, 709 (1994).

\bibitem{afterglow}
B.\ Paczy\'{n}ski and J.\ E.\ Rhoads, Astrophys.\ J.\ Lett.\ {\bf
418}, L5 (1993); P.\ M\'{e}sz\'{a}ros and M.\ J.\ Rees,  {\it
ibid.} {\bf 482}, L28 (1997); M.\ Vietri, {\it ibid.} {\bf 478},
L9 (1997); J.\ I.\ Katz, Astrophys.\ J.\ {\bf 422}, 248 (1994);
R.\ Sari and T.\ Piran, {\it ibid.} {\bf 485}, 270 (1997) .

\bibitem{observation}
E.\ Costa {\it et al.}, Nature {\bf 387}, 783 (1997); J.\ van
Paradijs {\it et al.}, {\it ibid.} {\bf 386}, 686 (1997); D.\ A.\
Frail {\it et al.}, {\it ibid.} {\bf 389}, 261 (1997).

\bibitem{dar}
For a dissenting view, see for instance A. Dar, Astrophys.\ J. Lett.\ {\bf
500}, L93 (1998).

\bibitem{redshift}
M.\ R.\ Metzger {\it et al.}, Nature {\bf 387}, 878 (1997); S.\
Kulkarni {\it et al.}, {\it ibid.} {\bf 393}, 35 (1998).

\bibitem{recent}
C. W. Akerlof {\it et al.}, GCN {\bf 205} (1999); Nature (1999), in
press; S.
Kulkarni {\it et al.}, Nature (1999), in press; astro-ph/9902272.

\bibitem{nsnsmerger}
S. I. Blinnikov, I. D. Novikov, T. V. Perevodchikova and A. G. Polnarev,
Soviet Astronomy Letters {\bf 10}, 177 (1984); D. Eichler, M. Livio, T.
Piran and D. N. Schramm, Nature {\bf 340}, 126 (1989);
see also H.-Th.\ Janka and M.\ Ruffert, Astro.\
Astrophys.\ Lett.\ {\bf 307}, L33 (1996); M.\ Ruffert, H.-Th.\
Janka, K.\ Takahashi and G.\ Sch\"{a}fer, Astro.\ Astrophys.\ {\bf
319}, 122 (1997).

\bibitem{variants}
See, for example, C.\ L.\ Fryer and S.\ E.\ Woosley, Astrophys.\
J.\ Lett. {\bf 502}, L9 (1998);  W.\ H.\ Lee and W.\ Klu\'{z}niak,
astro-ph/9903488.

\bibitem{collapses}
See, for example, G.\ M.\ Fuller and X.\ Shi, Astrophys.\ J.\
Lett.\ {\bf 502}, L5 (1998); A.\ Dar, B.\ Z.\ Kozlovsky, S.\
Nussinov and R.\ Ramaty, Astrophys.\ J.\ {\bf 388}, 164 (1992);
S.\ E.\ Woosley, {\it ibid.} {\bf 405}, 273 (1993); B.\
Paczy\'{n}ski, astro-ph/9706232; astro-ph/9712123.

\bibitem{berezinsky}
V. S. Berezinsky and O. F. Prilutsky, NASA Goddard Space Flight Center
Intern. Cosmic Ray Conf., Vol. 1 p29 - 32; Astro.\ Astrophys.\ {\bf 175},
309 (1987).

\bibitem{magnetic}
R. D. Blandford and R. L. Znajek, MNRAS {\bf 179}, 433 (1977).

\bibitem{superk}
SuperKamiokande Collaboration, Y.\ Fukada {\it et al.},
hep-ex/9803006; hep-ex/9805006; hep-ex/9807003.

\bibitem{kluzniak}
W.\ Klu\'{z}niak, Astrophys.\ J.\ Lett.\ {\bf 508}, L29 (1998).

\bibitem{ms}
S.\ P.\ Mikheyev and A.\ Yu.\ Smirnov, Nuovo Cimento C {\bf 9}, 17
(1986).

\bibitem{asymmetry}
R.\ Foot, M.\ J.\ Thomson and R.\ R.\ Volkas, Phys.\ Rev.\ D {\bf
53}, 5349 (1996); X. Shi, {\it ibid.} {\bf 54}, 2753 (1996); R.\
Foot and R.\ R.\ Volkas, {\it ibid.} {\bf 55}, 5147 (1997); {\it
ibid.} {\bf 56}, 6653 (1997) and Erratum {\it ibid.} {\bf 59},
029901 (1999); Astropart.\ Phys.\ {\bf 7}, 283 (1997); N. F. Bell,
R. Foot and R. R. Volkas, Phys.\ Rev.\ D {\bf 58}, 105010 (1998);
N. F. Bell, R. R. Volkas and Y. Y. Y. Wong, {\it ibid.} {\bf 59},
113001 (1999); R. Foot, Astropart.\ Phys.\ {\bf 10}, 253 (1999);
R. Foot and R. R. Volkas, hep-ph/9904336.

\bibitem{sn}
G.\ M.\ Fuller, R.\ Mayle, B.\ S.\ Meyer and J.\ R.\ Wilson,
Astrophys.\ J.\ {\bf 389}, 517 (1992); Y.-Z.\ Qian, G.\ M.\
Fuller, G.\ J.\ Mathews, R.\ W.\ Mayle, J.\ R.\ Wilson and S.\ E.\
Woosley, Phys.\ Rev.\ Lett.\ {\bf 71}, 1965 (1993); Y.-Z.\ Qian
and G.\ M.\ Fuller, Phys\ Rev.\ D {\bf 51}, 1479 (1995); {\it
ibid.} {\bf 52}, 656 (1995); G. Sigl, {\it ibid.} {\bf 51}, 4035
(1995).


\bibitem{juha}
H.\ Nunokawa, J.\ T.\ Peltoniemi, A.\ Rossi and J.\ W.\ F.\ Valle,
Phys.\ Rev.\ D {\bf 56}, 1704 (1997).

\bibitem{blinnikov}
S.\ Blinnikov, astro-ph/9902305.

\bibitem{wolfenstein}
L.\ Wolfenstein, Phys.\ Rev.\ D {\bf 17}, 2369 (1978); {\bf 20},
2634 (1979).

\bibitem{kuo}
For a review on matter effects on neutrino oscillations, see T.\
K.\ Kuo and J.\ Pantaleone, Rev.\ Mod.\ Phys.\ {\bf 61}, 937
(1989).

\bibitem{solar}
K.\ Lande (for the Homestake Collaboration), in {\it Neutrino'98},
Proceedings of the 18th International Conference on Neutrino
Physics and Astrophysics, Takayama, Japan, edited by Y.\ Suzuki
and Y.\ Totsuka, to be published in Nucl.\ Phys.\ B (Proc.\
Suppl.); GALLEX Collaboration, P.\ Anselmann {\it et al.}, Phys.\
Lett.\ B {\bf 342}, 440 (1995); W.\ Hampel {\it et al.}, {\it
ibid.} {\bf 338}, 364 (1996); SAGE Collaboration, V.\ Gavrin {\it
et al.}, in {\it Neutrino'98}; Y.\ Suzuki (for the SuperKamiokande
Collaboration), {\it ibid.}.

\bibitem{atmospheric}
See Ref.\ \protect\cite{superk} and T.\ Haines {\it et al.}, Phys.\
Rev.\ Lett.\ {\bf 57}, 1986 (1986); Kamiokande Collaboration, K.\
S.\ Hirata {\it et al.}, Phys.\ Lett.\ B {\bf 205}, 416 (1988);
{\bf 280}, 146 (1992); Y.\ Fukada {\it et al.}, {\it ibid.} {\bf
335}, 237 (1994); IMB Collaboration, D.\ Casper {\it et al.},
Phys.\ Rev.\ Lett.\ {\bf 66}, 2561 (1989); R.\ Becker-Szendy {\it
et al.}, Phys.\ Rev.\ D {\bf 46}, 3720 (1989); Soudan 2
Collaboration, W.\ W.\ M.\ Allison {\it et al.}, Phys.\ Lett.\ B
{\bf 391}, 491 (1997).

\bibitem{lsnd}
LSND Collaboration, C.\ Athanassopoulos {\it et al.}, Phys.\ Rev.\
Lett\ {\bf 75}, 2650 (1995); {\bf 77}, 3982 (1996);
nucl-ex/9706006.

\bibitem{pseudoDirac}See for example,
J. P. Bowes and R. R. Volkas, J. Phys.\ G {\bf 24}, 1249 (1998);
A. Geiser, Phys.\ Lett.\ B {\bf 444}, 358 (1999); P. Langacker,
Phys.\ Rev.\ D {\bf 58}, 093017 (1998); Y. Koide and H. Fusaoka,
Phys.\ Rev.\ D {\bf 59}, 053004 (1999); Z. Chacko and R. N.
Mohapatra, hep-ph/9905388; C. Giunti {\it et al.}, Phys.\ Rev.\ D
{\bf 46}, 3034 (1992); S. M. Bilenky and S. T. Petcov, Rev.\ Mod.\
Phys.\ {\bf 59}, 671 (1987).

\bibitem{wilson}
R.\ W.\ Mayle and J.\ R.\ Wilson, in {\it Supernovae}, Proceedings of the
Tenth Santa Cruz Summer Workshop, edited by S.\ E.\ Woosley,
(Springer Verlag, New York), p.333 (1991).

\bibitem{chooz}
CHOOZ Collaboration, M.\ Apollonio {\it et al.}, Phys.\ Lett.\ B
{\bf 420}, 397 (1998).

\bibitem{rate}
J.\ Goodman, A.\ Dar and S.\ Nussinov, Astrophys.\ J.\ Lett.\ {\bf
314}, L7 (1987).

\bibitem{epm}
R.\ Foot, H.\ Lew and R.\ R.\ Volkas, Phys.\ Lett.\ B {\bf 272},
67 (1991); see also the concluding section of: T.\ D.\ Lee and C.\
N.\ Yang, Phys.\ Rev.\ {\bf 104}, 254 (1956).

\bibitem{mohapatra}
Z. G. Berezhiani and R. N. Mohapatra, Phys.\ Rev.\ D {\bf 52}, 6607
(1995);
Z. G. Berezhiani, A. D. Dolgov and R. N. Mohapatra, Phys.\ Lett.\ B {\bf
375}, 26 (1996).

\bibitem{solution}
R.\ Foot and R.\ R.\ Volkas, Phys.\ Rev.\ D {\bf 52}, 6595 (1995);
R.\ Foot, H\ Lew and R.\ R.\ Volkas, Mod.\ Phys.\ Lett.\ A {\bf
7}, 2567 (1992); R.\ Foot, {\it ibid.} {\bf 9}, 169 (1994).


\end{thebibliography}
\end{document}